\documentclass[12pt]{article}

\usepackage{amsmath}
\usepackage{amssymb}
\usepackage{amsthm}

\usepackage{tikz,pgfplots}
\usetikzlibrary{plotmarks}

\usepackage{setspace}
\usepackage{natbib}

\usepackage{hyperref}

\usepackage[lined,algoruled,noline]{algorithm2e}

\newcommand{\mY}{\boldsymbol{Y}}
\newcommand{\my}{\boldsymbol{y}}
\newcommand{\mz}{\boldsymbol{z}}

\newcommand{\mT}{\boldsymbol{\Theta}}
\newcommand{\mt}{\boldsymbol{\theta}}
\newcommand{\mmu}{\boldsymbol{\mu}}
\newcommand{\muu}{\boldsymbol{u}}
\newcommand{\mX}{\boldsymbol{X}}
\newcommand{\mx}{\boldsymbol{x}}
\newcommand{\mbeta}{\boldsymbol{\beta}}
\newcommand{\mepsilon}{\boldsymbol{\epsilon}}
\newcommand{\mlambda}{\boldsymbol{\lambda}}
\newcommand{\mb}{\boldsymbol{b}}
\newcommand{\mB}{\boldsymbol{B}}
\newcommand{\mc}{\boldsymbol{c}}
\newcommand{\mxi}{\boldsymbol{\xi}}

\newcommand{\dd}{\mathrm{d}}
\newcommand{\normal}{\mathrm{N}}
\newcommand{\tsfrac}  [2] { {\textstyle \frac{#1}{#2} } }

\newcommand{\affaddr}[2]{\textsuperscript{#1}#2} % No op here. Customize it for different styles.
\newcommand{\affmark}[1]{\textsuperscript{#1}}
\newcommand{\email}[1]{\texttt{#1}}

\title{Sampling hyperparameters in hierarchical models:\\
improving on Gibbs for high-dimensional latent fields and large data sets.}

\author{
Richard A. Norton\affmark{1}, J. Andr\'es Christen\affmark{2} and Colin Fox\affmark{3}\\
\affaddr{1}{Department of Mathematics and Statistics,\\
University of Otago, Dunedin, New Zealand,\\
\email{richard.norton@otago.ac.nz},}\\
\affaddr{2}{Centro de Investigaci\'on en Matem\'aticas (CIMAT), CONACYT,\\
Guanajuato, GT, Mexico, \email{jac@cimat.mx}}\\
\affaddr{3}{Department of Physics,
University of Otago, \\Dunedin, New Zealand, \email{fox@physics@otago.ac.nz}.}
}

\date{20 October 2016}

\begin{document}

\maketitle

\begin{abstract}
We consider posterior sampling in the very common Bayesian hierarchical model in which observed data depends
on high-dimensional latent variables that, in turn, depend on relatively few hyperparameters. When the full conditional
over the latent variables has a known form, the marginal posterior distribution over hyperparameters is accessible and
can be sampled using a Markov chain Monte Carlo (MCMC) method on a low-dimensional parameter space. This may improve computational efficiency
over standard Gibbs sampling since computation is not over the high-dimensional space of latent variables and
correlations between hyperparameters and latent variables become irrelevant.
When the marginal posterior over hyperparameters depends on a fixed-dimensional sufficient statistic, precomputation
of the sufficient statistic renders the cost of the low-dimensional MCMC independent of data size. Then, when the
hyperparameters are the primary variables of interest, inference may be performed in big-data settings at modest cost.
Moreover, since the form of the full conditional for the latent variables does not depend on the form of the hyperprior
distribution, the method imposes no restriction on the hyperprior, unlike Gibbs sampling that typically requires conjugate distributions.
We demonstrate these efficiency gains in four computed examples.
\end{abstract}

\section{Introduction}

Suppose we have a hierarchical model structure in which the data model depends on a high-dimensional latent structure $\mX$, which in turn depends on a low-dimensional hyperparameter vector $\mT$.  That is, $f_{ \mY | \mX, \mT} ( \cdot | \mx, \mt )$ models the data $\mY$, and the joint prior distribution for $\mX$ and $\mT$ is defined by the hierarchical structure
$$
f_{ \mX, \mT }( \mx, \mt ) = f_{ \mX | \mT}( \mx | \mt ) f_{ \mT } (\mt).
$$
This is a very common hierarchical structure present in many statistical analyses \citep[for example][]{BCG,BardsleyRTO,GMI,PRS2007,WMNB2001}.  The posterior distribution is then
\begin{equation}
\label{eqn:post}
f_{ \mX, \mT | \mY }( \mx, \mt | \my) =
\frac{f_{ \mY | \mX, \mT} ( \my | \mx, \mt )f_{ \mX | \mT}( \mx | \mt ) f_{ \mT }(\mt)}
{f_{\mY}(\my)}.
\end{equation}
Indeed, in most relevant circumstances the normalizing constant
$$
f_{Y}(\my) =  \int \int f_{ \mY | \mX, \mT} ( \my | \mx, \mt ) f_{ \mX | \mT}( \mx | \mt ) f_{ \mT } (\mt ) \, \dd\mx \dd\mt
$$
is intractable and we must rely on Monte Carlo methods to explore the posterior distribution. 

A convenient, but restrictive, common method is to define conditional conjugate prior distribution such that \emph{all} of the full conditionals are of a known form and may be sampled from, and then progress by Gibbs sampling.  In the following, we require that the prior of the
latent structure $f_{ \mX | \mT}( \mx | \mt )$ has some sort of conjugancy to make the full conditional
$f_{ \mX | \mT, \mY } (\mx | \mt, \my)$ of a known form.  However, the prior for the hyperparameters $f_{ \mT } (\mt)$ may
have arbitrary form.  

Let us assume then that the full conditional for the parameters $\mX$
\begin{equation}
\label{eqn:FC_X}
f_{ \mX | \mT, \mY } (\mx | \mt, \my) =
\frac{f_{ \mY | \mX, \mT} ( \my | \mx, \mt )f_{ \mX | \mT}( \mx | \mt )}
{f_{ \mY| \mT}(\my|\mt)}
\end{equation}
belongs to a known family of distributions. This means that, in principle, by comparing this distribution with the numerator in
\eqref{eqn:post}, one can identify the $\mt$-dependence of the normalizing constant in the denominator of~\eqref{eqn:FC_X}, that is
\[
f_{ \mY | \mT }(\my | \mt) = \int f_{ \mY | \mX, \mT} ( \my | \mx', \mt )f_{ \mX | \mT}( \mx' | \mt ) \, \dd\mx' .
\]
Hence, if the full conditional for $\mX$ belongs to a known family of distributions, we should be able to determine
the marginal posterior distribution over hyperparameters
\begin{equation}
\label{eqn:marg_posta}
 f_{ \mT | \mY }(\mt | \my) = \frac{f_{\mY | \mT}(\my | \mt) f_{\mT}( \mt)}{ f_{ \mY}(\my)},
\end{equation}
excepting the normalizing constant $f_{Y}(\my)$.
This allows us to perform MCMC on the low-dimensional marginal posterior distribution over the hyperparameters,
for which automatic MCMC methods are available, and other customized MCMCs may be designed. We discuss automatic methods in Section~\ref{sec:lowdimmcmc}.

In many cases this marginal posterior distribution has a sufficient statistic of fixed low dimension that can be precomputed.
Consequently, the cost of evaluating $f_{\mT|\mY}(\mt|\my)$ at each iteration of an MCMC may grow not at all, or only slowly, with data size.  This improvement may become critical over the alternative of Gibbs sampling (or any other MCMC)
on the joint posterior distribution over $\mX,\mT|\mY$ since the cost of performing MCMC on the joint posterior distribution grows with the dimension of $\mX$ which, in many analyses, grows linearly with data size.   
Moreover, correlations between $\mX$ and $\mT$ that can adversely effect the computational efficiency of Gibbs sampling from the joint posterior distribution are irrelevant to MCMC on the marginal posterior distribution \citep{RueHeld}.

Note that once we have (pseudo) independent samples, after burn-in and thinning, of $\mT | \mY$, namely $\mt^{(i)}$ for $i=1,2, \ldots ,T$,
if inference on the latent structure is also required then we simply simulate
$\mx^{(i)}$ from $f_{ \mX | \mT, \mY } ( \cdot | \mt^{(i)}, \my)$ and
$(\mx^{(i)} , \mt^{(i)})$ for $i=1,2, \ldots ,T$ are (pseudo) \emph{independent} samples
from the joint posterior $f_{ \mX, \mT | \mY }( \cdot, \cdot | \my)$.  That is, the complete
results of the original Gibbs sampling are recovered.

However, performing MCMC on the marginal posterior distribution has at least three potential advantages:
the improved computational efficiency for `big data' just mentioned, an easier MCMC since correlations in a very
high-dimensional posterior are now avoided, and greater freedom of prior choice.

\cite{RueHeld} proposed a similar decomposition to give the \emph{one-block algorithm}.  From~\eqref{eqn:FC_X} the marginal posterior over hyperparameters is evaluated using the expression
\begin{equation}
\label{eqn:0a}
 f_{ \mY | \mT }(\my | \mt) = \frac{f_{\mY | \mX, \mT}(\my | \mx^*, \mt) f_{\mX | \mT}(\mx^* | \mt)}{ f_{\mX | \mY , \mT}(\mx^* | \my,\mt)}.
\end{equation}
with $\mx^*$ suitably chosen, restricted to $f_{\mX | \mY , \mT}(\mx^* | \my,\mt) > 0$.   This approach was proposed
in~\cite{RueHeld} and~\cite{SimpsonLindgrenRue2012} for linear Gaussian models.  \cite{Jupiter} in the same setting
canceled the terms in $\mx$ on the right-hand side of \eqref{eqn:0a} to give an expression that does not depend on $\mx$.

Indeed, as noted above, since $f_{\mX | \mY , \mT}(\mx^* | \my,\mt)$ has a known form, this cancellation appears to be always possible and the
selection of $\mx^*$ is unnecessary.  This is the procedure followed in the four examples we present in Section~\ref{sec:examples}.
In common with the one-block algorithm, the calculation that we use in~\eqref{eqn:marg_posta} does not require any special form of the hyperprior distribution $f_{ \mT}(\mt)$,
as opposed to the usual Gibbs sampling that requires conjugate distributions to be employed so that all full
conditionals are available. 

The approach we take is also related to partially collapsed Gibbs samplers \citep{Park2016,vanDyk2008,LIU1994} that perform strategic integrations and then apply Gibbs sampling on a reduced set of variables.  We are proposing an extreme version of partially collapsed Gibbs sampling by ``collapsing'' over \emph{all} latent variables and then sampling.
However, we do not then apply Gibbs sampling to the marginal posterior distribution, so we do not fit completely under the aegis of partially collapsed Gibbs samplers. Note also that we do not actually perform any integration, but use the algebraic route implied by \eqref{eqn:marg_posta}.

Our first example in Section~\ref{sec:examples} is the `pump failure data' analyzed by
\cite{GelfandSmith},
that sparked off the widespread use of Gibbs sampling.  Back then, analyzing a model with more than three parameters was in most cases prohibitive, and, consequently,  Bayesian statistics had a very marginal impact until that turning point.
The fact that \cite{GelfandSmith} showed a way to analyzing and sampling, in principle, from arbitrarily large models and provided an example with 11 parameters was revolutionary to most statisticians.
However, practically, Gibbs sampling may become arbitrarily inefficient with high correlations
\citep{Belisle1998} and certainly in latent structures with large data sets.
We now have a range of automatic, semiautomatic, and customized algorithms that we may employ
to deal with an arbitrary MCMC (with no conjugancy), over 10 to 15 parameters, to sample the hyperprior parameters, and
therefore the need and applicability of our approach.
 
The next section describes a measure of computational efficiency of MCMC methods, to achieve a quantitative
comparison of conventional Gibbs and our approach.  This is followed by four examples of sampling from the marginal posterior distribution to demonstrate its advantages over sampling from the joint posterior distribution.
We finish with some discussion.  Overly technical details are in the appendix.

\section{Computational considerations}
\subsection{Statistical and Computational Efficiency}

The statistical efficiency of (geometrically convergent) MCMC algorithms can be measured by the \emph{integrated autocorrelation time}  \citep[IACT, see][]{geyer92} for some real-valued statistic of interest, often one component of the state, or the log target density.
Let $h$ be the statistic of interest.  The IACT for $h$ is defined by
\begin{equation}
 \tau_{\mathrm{int}}^{(h)} = 1 + 2 \sum_{k=1}^\infty \rho_k
 \label{eq:tauint}
\end{equation}
where $\rho_k$ is the autocorrelation coefficient of $h$ at lag $k$. We estimate $\tau_{\mathrm{int}}^{(h)}$ using \emph{twice} the
IACT value computed using the (Matlab) function provided by
\cite{Wolff} since in the physics literature IACT is defined as $\tau_{\mathrm{int}}/2$ \citep[see also][]{Sokal}.
IACT has units of iterations, or steps, of the MCMC.  We prefer the statistics definition in~\eqref{eq:tauint} since it may be thought of as the length of the chain that has the same variance reducing power as one independent sample. That is, the lag required
in the chain to obtain a pseudo independent sample for the statistic $h$.

IACT suffers from a second more common ambiguity which is how one defines an `iteration', or step, of the MCMC. For example, when running a Gibbs sampler on a multivariate target distribution, one iteration could be the conditional sampling in one coordinate direction, or could be one sweep over all coordinates. For reversible Gibbs \citep[eg.][]{RobertsSahu, FoxParker2016} one iteration could be the forward then backwards sweeps over all coordinates. In each case the IACT depends on how one wishes to define an iteration. The ambiguity is further exacerbated by MCMC algorithms that have very unequal compute costs per iteration; for example, in the delayed acceptance of~\cite{amcmc}
the cost of a rejection is much less than the cost of an acceptance. There are examples of new sampling algorithms that are correctly reported as decreasing IACT, while the CPU effort required to achieve a given variance reduction is actually increased.

To measure the computational efficiency of MCMC methods we use \emph{computational cost per effective sample}
\citep[CCES, see][]{GMI}
$$
	\mathrm{CCES} = \frac{T}{N_{\mathrm{eff}}} 
$$
where $T$ is the compute time required to simulate the chain of length $N$ and $N_{\mathrm{eff}}$ is the \emph{effective sample size}
\citep[][p. 125]{GS1989, Liu} 
$$
	N_{\mathrm{eff}} = \frac{N}{\tau_{\mathrm{int}}^{(h)}} .
$$
Thus, CCES measures the compute time required to reduce variance in estimates of $h$ by the same amount as one independent sample. In our simulations, $h$ is usually one component of the hyperparameter $\mT$.

\subsection{Tuning-free MCMCs for low-dimensional distributions\label{sec:lowdimmcmc}}

When there are a small number of hyperparameters (say, fewer than 15) the task of sampling from the marginal posterior distribution over hyperparameters represents a case of MCMC sampling over a low-dimensional space.
There are now several excellent tuning-free algorithms for sampling from continuous distributions that are efficient in low-dimensional problems with no special form, for which computer codes are readily available. We mention here some of these,
that may be useful for our hyperparameter sampling problem. 

Independent doubly adaptive rejection Metropolis sampling (IDARM), available in the IA2RMS Matlab package \url{http://a2rms.sourceforge.net/}, 
is a development of the adaptive rejection sampler (ARS), for sampling from univariate distributions. It can be thought of as an independence Metropolis algorithm with the proposal distribution tending to the target distribution, almost everywhere, as iterations progress. Thus, asymptotic in sample size, IDARM produces a chain of independent samples from the target distribution at the theoretical cost of one evaluation of the density function per independent sample. Hence, IDARM is particularly efficient for drawing many samples from a univariate distribution of unchanging form, as occurs in the examples that have a single hyperparameter.   However, this package is currently functional, but not robust.

The adaptive Metropolis (AM) algorithm is a random-walk Metropolis (RWM) algorithm that uses a multivariate normal proposal, with the covariance being primarily a scaled version of the sample covariance over the existing chain. The scaling may be chosen to be optimal in certain settings.
The resulting algorithm is asymptotically ergodic, and thus AM is suitable for automatic sampling from general multivariate distributions. 
A \textit{delayed-rejection} version of AM may be found in the Matlab package DRAM (see~\url{http://helios.fmi.fi/~lainema/dram/}).

The t-walk algorithm~\citep[t-walk, see ][]{twalk} also enables automatic sampling from multivariate distributions over a continuous state space. The t-walk algorithm maintains a pair of states and is strictly a non-adaptive Metropolis-Hastings algorithm on the product space, though shows adaptive properties on the original space and provides effective tuning-free sampling in many settings.  It is available in R, Python, C and Matlab, \url{http://www.cimat.mx/~jac/twalk/}.

A measure of the efficiency of these sampling algorithms may be gained by calculating the IACT when sampling from a $d$-dimensional normal distribution with zero mean. We considered IACT for one component of the state variable, and for the log target density function, which gave similar results for these algorithms. Since AM and t-walk are both close to invariant to affine transformations, the structure of the covariance matrix is not crucial for these two algorithms.  Asymptotic in the chain length,  AM has $\tau_{\mathrm{int}}\approx 5d$, while t-walk has $\tau_{\mathrm{int}}\approx 12d$ for $d\geq 3$~\citep{twalk}.
However, even when initializing with a state in the support of the target distribution, AM can suffer from a long burn-in period when convergence of the proposal covariance is slow. For example, for $d=10$, overestimating the initial proposal covariance by a factor of $2$ in AM leads to a burn-in length of around $10^4$. The delayed rejection variant of AM (DRAM) improves this dramatically, though compute cost per iteration is roughly doubled, and burn-in is around $10^3$ when the proposal covariance is either over- or under-estimated. For the same compute cost as this DRAM burn-in, t-walk will have generated about 160 effectively independent samples, which may be sufficient for some purposes.  IDARM may also be applied within Gibbs sampling to draw from multivariate distributions.  \cite{FoxParker2016} showed that the (distributional) convergence of the Gibbs sampler is exactly the same as for Gauss-Seidel iteration applied to the precision matrix, so convergence may be determined using standard results from numerical methods. For example, when sampling from a bivariate normal distribution with correlation $\rho$ the asymptotic average reduction factor~\citep[called the \emph{convergence rate} in statistics, see][]{RobertCasella} per sweep is exactly $\rho^2$, and hence is both problem and parametrization specific. Since the implementation of IDARM in IA2RMS has a burn-in of order $10$ function evaluations, the cost of a sweep is $\approx 10 d$ and IACT can be estimated by $10 d (1+\rho^2)/(1-\rho^2)$, which is greater than for either AM or t-walk. 

Accordingly, for tuning-free automatic sampling, we use IDARM for sampling univariate distributions, and either AM or t-walk for multivariate marginal posterior distributions.

\section{Examples\label{sec:examples}}
\subsection{Pump failures}

This example, originally presented in \cite{Gaver1987}, has been used as an example in several papers on analysis of the MCMC
\citep[see][p. 385 and references therein]{RobertCasella} and was presented as an example for the Gibbs sampler in the original
\cite{GelfandSmith} paper.  We present it here to demonstrate the feasibility of working with the marginalized posterior distribution.

Running times $t_i$ and number of failures $p_i$ are recorded for pumps $i=1,\dotsc,n$.  The hierarchical model structure is 
$$
p_i \sim \operatorname{Po}(t_i \lambda_i),~~ \lambda_i \sim \operatorname{Ga}( \alpha, \beta) ~~\text{and}~~
\beta \sim \operatorname{Ga}(\gamma, \delta)
$$
for known positive parameters $\alpha$, $\beta$, $\gamma$ and $\delta$.  We use the convention that $\beta$ and $\delta$ are \emph{rate} parameters of the gamma distribution.  It is easily seen that 
\begin{align*}
	\lambda_i | \beta,\mathbf{p} &\sim \operatorname{Ga}( p_i + \alpha, t_i + \beta) \quad \mbox{for $i=1,\dotsc,n$},\\
	\beta | \mlambda, \mathbf{p} &\sim \operatorname{Ga}\left( n\alpha + \gamma, \delta + \sum_{i=1}^n \lambda_i \right), 
\end{align*}
and hence, Gibbs sampling is possible for this example.  

Identifying $\mathbf{p}=(p_1,p_2,\dotsc,p_n)$ with $\my$, $\mlambda = (\lambda_1,\lambda_2,\dotsc,\lambda_n)$ with $\mx$ and $\mt$ with $\beta$ we see that $f_{\mX|\mT,\mY}(\mx|\mt,\my)$ is a product of gamma distributions and from~\eqref{eqn:marg_posta} it is easily shown that the marginal posterior distribution over $\beta$ satisfies 
$$
f_{B|\mathbf{P}} (\beta|\mathbf{p}) 
\propto \beta^{n\alpha + \gamma -1} e^{-\delta \beta} \prod_{i=1}^n (\beta + t_i)^{-(\alpha + p_i)}.
$$ 
This one-dimensional distribution remains one dimensional as sample size increases, in contrast to the joint posterior distribution that increases in dimension as sample size increases.

However, since there is no fixed-dimension sufficient statistic, the cost of evaluating $f_{B|\mathbf{P}}(\beta|\mathbf{p})$ will increase (na\"ive evaluation increases linearly) with sample size, so we still expect computational efficiency of MCMC to remain dependent on sample size.

Nevertheless, we can exploit the fact that the marginal posterior distribution is one-dimensional to improve efficiency, by employing the IDARMS sampler that constructs an independent proposal for the Metropolis-Hastings algorithm that converges to
$f_{B|\mathbf{P}} (\beta|\mathbf{p})$.  
We used the original data presented in Table \ref{table1} and parameters $\alpha = 1.8$, $\gamma = 0.01$ and $\delta = 1$.  

\begin{table}
\begin{center}
\begin{tabular}{|l|c|c|c|c|c|c|c|c|c|c|}
\hline
Pump ($i$)               & 1     & 2     & 3     & 4      & 5    & 6     & 7    & 8    & 9   & 10    \\ \hline
Run time ($t_i$)         & 94.32 & 15.72 & 62.88 & 125.76 & 5.24 & 31.44 & 1.05 & 1.05 & 2.1 & 10.48 \\
\# of failures ($p_i$) & 5     & 1     & 5     & 14     & 3    & 19    & 1    & 1    & 4   & 22 \\ \hline
\end{tabular}
\end{center}
\caption{Data on pump failures from \cite[p. 385]{RobertCasella}.}
\label{table1}
\end{table}

Results comparing IDARMS targeting $f_{B|\mathbf{P}} (\beta|\mathbf{p})$ and Gibbs sampling over the joint posterior distribution are given in Table \ref{table0}.  Even for this small problem, sampling from the marginal distribution for $\beta$ was almost an order of magnitude faster than Gibbs sampling from the full posterior distribution including latent variables $\lambda_i$.  This results from the sampler over the marginal posterior distribution having both smaller IACT and smaller compute time per sample.
For problems with larger $n$, we expect that IDARMS will still beat Gibbs sampling even though the cost of both algorithms grows linearly with $n$.  
Moreover, it is not guaranteed that, with a hypothetical larger sample size, the Gibbs sampler will continue to perform well.  That is, the IACT for the Gibbs sampler divided by the dimension will need to remain roughly constant, which only happens in optimal cases~\citep{roberts01}. 

\begin{table}
\begin{center}
\begin{tabular}{|l|r|l|l|r|l|l|}
\hline
                & \multicolumn{3}{c|}{Gibbs}  & \multicolumn{3}{c|}{IDARMS} \\ \hline
burn            & \multicolumn{3}{r|}{100}    & \multicolumn{3}{r|}{100}                 \\ \hline
samples         & \multicolumn{3}{r|}{10000}  & \multicolumn{3}{r|}{10000}               \\ \hline
acceptance rate & \multicolumn{3}{r|}{100\%}  & \multicolumn{3}{r|}{97\%}               \\ \hline
time            & \multicolumn{3}{r|}{0.62}    & \multicolumn{3}{r|}{0.25}                \\ \hline
mean of $\beta$ & \multicolumn{3}{r|}{2.464}  & \multicolumn{3}{r|}{2.467}               \\ \hline
IACT            & \multicolumn{3}{r|}{1.9}    & \multicolumn{3}{r|}{1.0}                 \\ \hline
CCES            & \multicolumn{3}{r|}{1.1e-4} & \multicolumn{3}{r|}{2.8e-5}                \\ \hline
\end{tabular}
\caption{Pump failures; results of Gibbs sampling on $\mlambda,\beta|\mathbf{p}$ and IDARMS for sampling $\beta|\mathbf{p}$.}
\label{table0}
\end{center}
\end{table}

\subsection{Censored data}

A classical example used to illustrate the Gibbs sampler is using latent variables in Gaussian censored data. 
Suppose $y_i$ is observed with right censoring, that is, if $y_i > a$ then only an ``observation above $a$'' is recorded.  Let $y_1 < y_2 < \cdots < y_m$ be the uncensored observations and let there be $n-m$ additional censored observations.  It is assumed that observations are normally distributed with unknown mean $\mu$ and precision $\lambda$.  It is further assumed that $\mu | \lambda \sim \normal( \mu_0, (k_0 \lambda)^{-1})$ and $\lambda \sim \operatorname{Ga}( \alpha, \beta)$, with $k_0=\alpha=1$ and $\beta = 0.1$.  Here $(k_0 \lambda)^{-1}$ is the variance of the normal distribution and $\beta$ is the rate of the gamma distribution.  It follows from~\eqref{eqn:marg_posta} that
\begin{equation}
\label{eq:post}
f_{M,\Lambda|\mY}(\mu, \lambda | \my ) \propto \lambda^{\alpha_1 -1/2} \exp\left(-\lambda \left( \frac{k_1}{2}(\mu-\mu_1)^2 + \beta_1 \right)\right)
\left( 1 - \Phi(\sqrt{\lambda}(a - \mu)) \right)^{n-m} 
\end{equation}
for parameters $\alpha_1 = \alpha + \frac{m}{2}$, $k_1 = k_0 + m$, 
$\mu_1 = \frac{1}{k_1}(k_0 \mu_0 + \sum_{i=1}^m y_i)$, and 
$\beta_1 = \beta + \frac{1}{2}( k_0 \mu_0^2 - k_1 \mu_1^2 + \sum_{i=1}^m y_i^2)$
dependent on the uncensored data.  This is a two-dimensional posterior distribution, with sufficient statistics $m$, $\sum_{i=1}^m y_i$, and $\sum_{i=1}^m y_i^2$.  Hence, once these statistics have been computed, the cost of sampling from the marginal posterior distribution is independent of sample size whatever MCMC method we use.  Moreover, we have seen that IACT for the integrated case remains constant for increasing sample size and hence CCES is also independent of sample size.

The conventional approach to this problem is to add latent variables $x_i$ for the unobserved data, that is $x_i = y_{m+i}$, and use block-Gibbs sampling.  The full conditional of $\mu, \lambda | \my, \mx$ is a normal-gamma distribution and each $x_i |  \mu, \lambda , \my$ full conditional is an independent and identically distributed truncated normal distribution.  This creates a block Gibbs sampler with increasing dimension as the sample size increases.  A sweep of block-Gibbs sampling is
\begin{align*}
	&\mbox{sample $\mu,\lambda|\my,\mx$ by sampling $\lambda | \my,\mx \sim \operatorname{Ga}( \alpha_2,\beta_2)$ then $\mu | \lambda, \my, \mx  \sim \normal( \mu_2, (k_2 \lambda)^{-1})$}, \\
	&\mbox{sample $x_i | \mu, \lambda, \my \sim \normal(\mu,\lambda^{-1})$ truncated to $x_i \in [a,\infty)$ for $i=1,\dotsc,n-m$}.
\end{align*}
where $\alpha_2 = \alpha_0 + \frac{n}{2}$, $k_2 = k_0 + n$, $\mu_2 = \frac{1}{k_2} (k_0 \mu_0 + \sum_{i=1}^m y_i + \sum_{i=1}^{n-m} x_i)$, and $\beta_2 = \beta + \frac{1}{2}( k_0 \mu_0^2 - k_2 \mu_2^2 + \sum_{i=1}^m y_i^2 + \sum_{i=1}^{n-m} x_i^2)$.

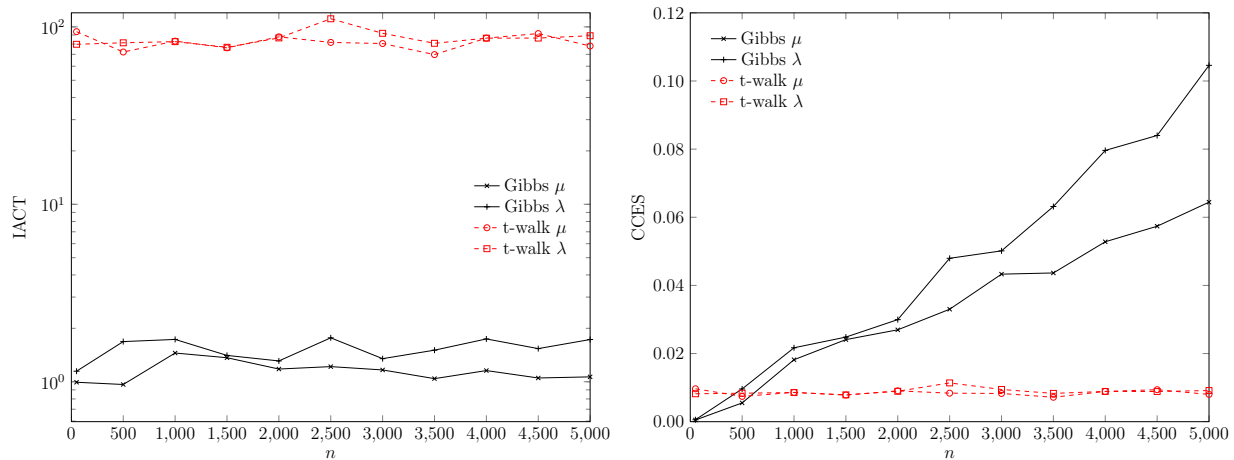
\begin{figure}
\begin{center}
\resizebox{0.49\textwidth}{!}{% This file was created by matlab2tikz v0.2.3.
% Copyright (c) 2008--2012, Nico Schlömer <nico.schloemer@gmail.com>
% All rights reserved.
% 
% The latest updates can be retrieved from
%   http://www.mathworks.com/matlabcentral/fileexchange/22022-matlab2tikz
% where you can also make suggestions and rate matlab2tikz.
% 
% 
% 
\begin{tikzpicture}

\begin{semilogyaxis}[%
view={0}{90},
width=4.82222222222222in,
height=3.80333333333333in,
scale only axis,
xmin=0, xmax=5000,
xlabel={$n$},
ymin=0, ymax=120,
ylabel={IACT},
legend style={at={(0.97,0.5)},anchor=east,fill=none,draw=none,align=left}]
\addplot [
color=black,
solid,
mark=x,
mark options={solid}
]
coordinates{
 (50,0.993272955543579)(500,0.96493367372247)(1000,1.45151913216141)(1500,1.36552234349092)(2000,1.17921576427224)(2500,1.21684818269601)(3000,1.16535210898397)(3500,1.04103757446023)(4000,1.15590216922537)(4500,1.04994978765954)(5000,1.06533019430244) 
};
\addlegendentry{Gibbs $\mu$};

\addplot [
color=black,
solid,
mark=+,
mark options={solid}
]
coordinates{
 (50,1.14636202201044)(500,1.68386461855207)(1000,1.73161972750429)(1500,1.40723862105083)(2000,1.30988084003242)(2500,1.76812061059593)(3000,1.34924955157946)(3500,1.50619696014788)(4000,1.74270470584994)(4500,1.53735477078485)(5000,1.72944355200453) 
};
\addlegendentry{Gibbs $\lambda$};

\addplot [
color=red,
dashed,
mark=o,
mark options={solid}
]
coordinates{
 (50,93.9586898856007)(500,72.1620361067241)(1000,83.1122539074681)(1500,76.1948404829352)(2000,87.869619491574)(2500,81.7235440538049)(3000,80.5525890037218)(3500,69.7737674264038)(4000,86.4157643104938)(4500,91.6723079504117)(5000,78.0113458517815) 
};
\addlegendentry{t-walk $\mu$};

\addplot [
color=red,
dashed,
mark=square,
mark options={solid}
]
coordinates{
 (50,79.7673761464189)(500,81.3284681496084)(1000,82.5619555887596)(1500,76.743064464257)(2000,86.6427128164318)(2500,111.08151590055)(3000,91.9912210177004)(3500,80.8008882009494)(4000,86.3925998873988)(4500,86.4655944525643)(5000,89.1330336912904) 
};
\addlegendentry{t-walk $\lambda$};

\end{semilogyaxis}
\end{tikzpicture}%
}
\resizebox{0.49\textwidth}{!}{% This file was created by matlab2tikz v0.2.3.
% Copyright (c) 2008--2012, Nico Schlömer <nico.schloemer@gmail.com>
% All rights reserved.
% 
% The latest updates can be retrieved from
%   http://www.mathworks.com/matlabcentral/fileexchange/22022-matlab2tikz
% where you can also make suggestions and rate matlab2tikz.
% 
% 
% 
\begin{tikzpicture}

\begin{axis}[%
view={0}{90},
width=4.82222222222222in,
height=3.80333333333333in,
scale only axis,
xmin=0, xmax=5000,
xlabel={$n$},
ymin=0, 
ymax=0.12,
y tick label style={
        /pgf/number format/.cd,
        fixed,
        fixed zerofill,
        precision=2,
        /tikz/.cd
    },
ylabel={CCES},
legend style={at={(0.03,0.97)},anchor=north west,fill=none,draw=none,align=left}]
\addplot [
color=black,
solid,
mark=x,
mark options={solid}
]
coordinates{
 (50,0.000497382907646302)(500,0.00550878532492433)(1000,0.0181855370907411)(1500,0.0240932640605329)(2000,0.0269684356909954)(2500,0.0330043997680379)(3000,0.0433050358929215)(3500,0.0436562405186766)(4000,0.0528089768192853)(4500,0.0573712284321203)(5000,0.0644282927051048) 
};
\addlegendentry{Gibbs $\mu$};

\addplot [
color=black,
solid,
mark=+,
mark options={solid}
]
coordinates{
 (50,0.000574042485039581)(500,0.00961314642907458)(1000,0.0216948120654098)(1500,0.0248293056900707)(2000,0.0299567206168439)(2500,0.0479564832327102)(3000,0.0501387518924211)(3500,0.0631628467347194)(4000,0.079617855960737)(4500,0.0840039521628129)(5000,0.104592074815323) 
};
\addlegendentry{Gibbs $\lambda$};

\addplot [
color=red,
dashed,
mark=o,
mark options={solid}
]
coordinates{
 (50,0.00965452237727019)(500,0.0074161724719558)(1000,0.00860881416467092)(1500,0.00781899156244794)(2000,0.00904942241260511)(2500,0.00837356909885697)(3000,0.00827849644366424)(3500,0.00717057063303276)(4000,0.00888439788394885)(4500,0.00939095297818439)(5000,0.00800538759835718) 
};
\addlegendentry{t-walk $\mu$};

\addplot [
color=red,
dashed,
mark=square,
mark options={solid}
]
coordinates{
 (50,0.00819632456475696)(500,0.00835821685775934)(1000,0.00855181395425469)(1500,0.00787524942265366)(2000,0.00892306705988537)(2500,0.01138164967964)(3000,0.00945405983174765)(3500,0.00830381527940025)(4000,0.00888201635144526)(4500,0.00885757487609044)(5000,0.00914667571396494) 
};
\addlegendentry{t-walk $\lambda$};

\end{axis}
\end{tikzpicture}%
}
\end{center}
\caption{Censored data; IACT and CCES for increasing $n$.  Here and in all other figures, `Gibbs $\mu$' etc. means
taking parameter $\mu$ as the $h$ scalar function for calculation of the IACT and CCES.}
\label{fig2a}
\end{figure}

We generated synthetic data using $\mu = 2$, $\lambda = 1$ and $a = 3$, and compared block-Gibbs with t-walk for sampling from \eqref{eq:post}.  In Figure~\ref{fig2a} we see that IACT remains approximately constant for increasing sample size so the computational efficiency of each method scales in the same way as computation time; linearly for block-Gibbs and constant for t-walk.  Burn in was $10$ times IACT and chain length was $1000$ times IACT.  

The AM algorithm will give better performance than t-walk for sampling this marginal posterior distribution, though one can see from the CCES in Figure~\ref{fig2a} that this is a mute point since t-walk already provides fast sampling at all sample sizes.
In this case, the Gibbs sampler may become arbitrarily inefficient, in comparison to the marginalized alternative.

\subsection{Dyes: Variance component model}
\label{sec:dyes}

\cite{BoxTiao} analyzed data regarding batch variations in yields of dyestuff.  The OpenBugs software \citep{openbugs2009} also uses this example.  Let $y_{ij}$ be the yield of sample $j$ from batch $i$.  The model is $y_{ij}|\mu_i,t_w \sim \normal(\mu_i, t_w^{-1} )$ for $i=1,\dotsc,B$ and $j=1,\dotsc,S$ where $\mu_i |\theta,t_b \sim \normal(\theta,t_b^{-1})$.  We are trying to infer the hyperparameters $\theta$, $t_w$ and $t_b$ from data.

From~\eqref{eqn:marg_posta} the posterior distribution marginalized over $\mmu = (\mu_1,\dotsc,\mu_B)$ satisfies
\begin{multline}
\label{eq:f1}
	f_{\Theta,T_w,T_b|\mY}(\theta,t_w,t_b | \my) \propto \left( \frac{t_w^S t_b}{t_b + S t_w}\right)^{B/2} \times \\ \exp \left( \frac{B}{2(t_b + S t_w)} \left( S^2 t_w^2 r_1 + 2 S t_w t_b \theta r_2 + t_b^2 \theta^2 \right) - \frac{BS}{2} t_w r_3 - \frac{B}{2} t_b \theta^2 \right) f_\Theta(\theta) f_{T_w}(t_w) f_{T_b}(t_b)
\end{multline}
where $f_\Theta(\theta)$, $f_{T_w}(t_w)$ and $f_{T_b}(t_b)$ are hyperprior distributions and 
$$
	r_1 = \frac{1}{B} \sum_{i=1}^B \left( \frac{1}{S} \sum_{j=1}^S y_{ij} \right)^2, \quad r_2 = \frac{1}{BS} \sum_{i=1}^B \sum_{j=1}^S y_{ij}, \quad \mbox{and} \quad r_3 = \frac{1}{BS} \sum_{i=1}^B \sum_{j=1}^S y_{ij}^2.
$$
Notice that there is so far no restriction on how we choose the hyperprior distributions and it is not necessary for them to be conjugate priors.  Moreover, the complexity of sampling from this distribution does not increase with $B$ nor $S$ since we may
precompute $r_1$, $r_2$, and $r_3$. In particular, the cost of evaluating the marginal posterior distribution in this example
is independent of sample size. Therefore t-walk or DRAM, or any other MCMC method, will have computational cost that is independent
of $B$ and $S$.  

However, the likelihood part of \eqref{eq:f1} does not decay as $t_b \rightarrow \infty$, so, depending on the choice of prior for $T_b$, \eqref{eq:f1} may have a heavy tail in the $t_b$ coordinate direction.  Thus, MCMC methods utilizing local moves, such as t-walk and RWM, including AM, could perform poorly even though the cost per iteration is independent of $B$ and $S$.

Instead of working directly with \eqref{eq:f1} we perform a simplifying coordinate transformation $(t_w,t_b) \in [0,\infty)^2 \leftrightarrow (x,w) \in [0,\infty)\times[0,1)$ defined by 
$$
	x = S t_w \qquad \mbox{and} \qquad w = \frac{t_b}{t_b + St_w}.
$$
Then \eqref{eq:f1} becomes
\begin{multline*}
	f_{\Theta,X,W|\mY}(\theta,x,w|\my) \propto 
	\frac{x^{BS/2} w^{B/2}}{(1-w)^2} \exp \left( - \frac{B}{2} \left[ R_3 x + \left( (\theta-r_2)^2 + R_2\right) xw  \right] \right) \times \\ 
	f_{\Theta}(\theta) f_{T_w}\left( \tsfrac{x}{S} \right) f_{T_B}\left( \tsfrac{xw}{1-w} \right)
\end{multline*}
where $R_2 = r_1 - r_2^2$ and $R_3 = r_3 - r_1$.  Further details of the coordinate transformation are left to the appendix.

After hyperprior distributions are chosen, we apply a tailor-made MCMC method to sample from $f_{\Theta,X,W|\mY}(\theta,x,w|\my)$.  The effect of the coordinate transformation and hyperprior choice is that we can extract two full conditional distributions from known families of distributions from $f_{\Theta,X,W|\mY}(\theta,x,w|\my)$, so we choose our tailor-made MCMC method to be a Metropolis within Gibbs (MwG) method.

\subsubsection[Tailor-made MCMC to sample from marginal posterior]{Tailor-made MCMC to sample from $f_{\Theta,X,W|\mY}(\theta,x,w|\my)$}
\label{sec:dyesmwg}

To apply MwG to $f_{\Theta,X,W|\mY}(\theta,x,w|\my)$ we choose the same conjugate hyperpriors as for Gibbs sampling, 
$$
	\theta \sim \normal(m,\lambda^{-1}), \quad t_w \sim \operatorname{Ga}(a,b), \quad \mbox{and} \quad t_b \sim \operatorname{Ga}(c,d),
$$
with $m=0$, $\lambda = 10^{-10}$, and $a=b=c=d=10^{-3}$.  This choice of hyperpriors allows two of the three full conditional distributions of $f_{\Theta,X,W|\mY}(\theta,x,w|\my)$ to be from known families of distributions, 
\begin{align*}
	\theta | x,w,\my &\sim \normal\left( r_2 + \frac{\lambda(m-r_2)}{B x w + \lambda}, (B x w + \lambda)^{-1} \right) \\
	x | \theta,w,\my &\sim \operatorname{Ga}\left( \frac{BS}{2} + a + c, \frac{BR_3}{2} + \frac{B}{2}w ((\theta-r_2)^2 + R_2) + \frac{b}{S} + d \frac{w}{1-w} \right).
\end{align*}
The remaining conditional distribution satisfies 
$$
	f_{W|\Theta,X,\mY}(w|\theta,x,\my) \propto f(w) := w^{b_1} \left( \frac{1}{1-w} \right)^{b_2} \exp\left( -b_3 w - d x \frac{w}{1-w} \right)
$$	
where $b_1 = B/2 + c - 1$, $b_2 = c+1$, and $b_3 = B x ((\theta-r_2)^2 + R_2)/2$.  

To sample from $f_{W|\Theta,X,\mY}(w|\theta,x,\my)$ we perform $5$ iterations of RWM with proposal $w' \sim \normal(w,u^2)$ for some $u>0$.  If $w' \notin [0,1)$ then $w'$ is rejected.  We used $5$ iterations of RWM to make the cost of drawing from $f_{W|\Theta,X,\mY}(w|\theta,x,\my)$ comparable to drawing samples from $f_{\Theta|X,W,\mY}(\theta|x,w,\my)$ and $f_{X|\Theta,W,\mY}(w|\theta,w,\my)$, and $u$ was chosen to be twice the standard deviation of $w$ values after a training run of length twice burn in.

In this case t-walk does not perform sufficiently well and therefore the need for the
above tailor-made MwG algorithm.  However, t-walk requires very little effort to use and so is included in this
example for comparison.

\subsubsection{Simulations}

The more conventional approach to sampling for this problem is to perform Gibbs sampling on $f_{\mathbf{M},\Theta,T_w,T_b|\mY}(\mmu,\theta,t_w,t_b|\my)$, which includes latent variables $\mmu$, with conjugate hyperprior distributions.
Then
\begin{align*}
	\mu_i | \mmu_{-i}, \theta, t_w, t_b,\theta &\sim \normal\left(\frac{v_{1i}}{v},v^{-1}\right) \qquad \mbox{for $i=1,\dotsc,B$},\\
	\theta | \mmu, t_w, t_b, \my &\sim \normal\left( \bar{\mu} + \frac{ \lambda(m-\bar{\mu})}{B t_b + \lambda}, (B t_b + \lambda)^{-1} \right), \\
	t_w | \mmu, \theta, t_b, \my &\sim \operatorname{Ga}\left( \frac{BS}{2} + a, \frac{S}{2} |\mmu|^2 - S \mathbf{z}_1 \cdot \mmu + \frac{BS}{2} r_3 + b \right), \\
	t_b | \mmu, \theta, t_w, \my &\sim \operatorname{Ga}\left( \frac{B}{2} + c , \frac{1}{2} | \mmu |^2 - B\theta \bar{\mu} + \frac{B}{2} \theta^2 + d  \right),
\end{align*}
where $\mmu_{-i} = (\mu_1,\dotsc,\mu_{i-1},\mu_{i},\dotsc,\mu_B)$, $\bar{\mu} = \frac{1}{B}\sum_{i=1}^B \mu_i$, $v = S t_w + t_b$, $v_{1i} = S t_w z_{1i} + t_b \theta$, and $\mathbf{z}_1 = (z_{11},z_{12},\dotsc,z_{1B})$ with $z_{1i} = \frac{1}{S} \sum_{j=1}^S y_{ij}$.

We expect the cost of Gibbs sampling from $f_{\mathbf{M},\Theta,T_w,T_b|\mY}(\mmu,\theta,t_w,t_b|\my)$ to increase at least linearly with $B$ since $\mmu$ has $B$ dimensions.

To compare these MCMC methods we performed simulations using data in Table \ref{table2} and initial values $\theta = 1500$ and $t_w = t_b = 1$.

\begin{table}[h]
\begin{center}
\begin{tabular}{|l|lllll|}
\hline
Batch & \multicolumn{5}{l|}{Yield in grams}        \\ \hline
1     & 1545 & 1440 & 1440 & 1520 & 1580 \\
2     & 1540 & 1555 & 1490 & 1560 & 1495 \\
3     & 1595 & 1550 & 1605 & 1510 & 1560 \\
4     & 1445 & 1440 & 1595 & 1465 & 1545 \\
5     & 1595 & 1630 & 1515 & 1635 & 1625 \\
6     & 1520 & 1455 & 1450 & 1480 & 1445 \\ \hline
\end{tabular}
\end{center}
\caption{Data on yields from dyestuff.}
\label{table2}
\end{table}

The comparison between Gibbs on $f_{\mathbf{M},\Theta,T_w,T_b|\mY}(\mmu,\theta,t_w,t_b|\my)$, our tailor-made MCMC, and t-walk on $f_{\Theta,X,W|\mY}(\theta,x,w|\my)$ is summarized in Table \ref{table3} for the parameters $\theta$, $s_w = 1/t_w$ and $s_b = 1/t_b$.
We found that our tailor-made MCMC on $f_{\Theta,X,W|\mY}(\theta,x,w|\my)$ outperforms (in both IACT and compute time) Gibbs sampling from $f_{\mathbf{M},\Theta,T_w,T_b|\mY}(\mmu,\theta,t_w,t_b|\my)$, but t-walk is not competitive when data size is small.
\begin{table}[h]
\begin{center}
\begin{tabular}{|l|r|r|r||r|r|r||r|r|r|}
\hline
  & \multicolumn{3}{c|}{Gibbs} & \multicolumn{3}{c|}{Tailor-made MwG} & \multicolumn{3}{c|}{t-walk} \\ \hline
burn & \multicolumn{3}{r|}{1e4} & \multicolumn{3}{r|}{1e4} & \multicolumn{3}{r|}{1e5} \\ \hline
samples & \multicolumn{3}{r|}{1e5} & \multicolumn{3}{r|}{1e5} & \multicolumn{3}{r|}{1e6} \\ \hline
CPU time (sec) & \multicolumn{3}{r|}{15.0} & \multicolumn{3}{r|}{5.8} & \multicolumn{3}{r|}{55.0} \\ \hline
statistic & \multicolumn{1}{c|}{$\theta$} & \multicolumn{1}{c|}{$s_w$} & \multicolumn{1}{c|}{$s_b$} & \multicolumn{1}{c|}{$\theta$} & \multicolumn{1}{c|}{$s_w$} & \multicolumn{1}{c|}{$s_b$} & \multicolumn{1}{c|}{$\theta$} & \multicolumn{1}{c|}{$s_w$} & \multicolumn{1}{c|}{$s_b$} \\ \hline
mean & 1527 & 3002 & 2264 & 1527 & 3019 & 2240 & 1522 & 2866 & 2576 \\ \hline
IACT & 3.0 & 29 & 4.2 & 1.0 & 14 & 4.2 & 420 & 1200 & 200 \\ \hline
CCES (sec) & 4e-4 & 4e-3 & 6e-4 & 5e-5 & 7e-4 & 2e-4 & 2e-2 & 6e-2 & 1e-2 \\ \hline
\end{tabular}
\end{center}
\caption{Results of Gibbs sampling from $f_{\mathbf{M},\Theta,T_w,T_b|\mY}(\mmu,\theta,t_w,t_b|\my)$, our tailor-made MCMC method, and t-walk for sampling from $f_{\Theta,X,W|\mY}(\theta,x,w|\my)$ for the Dyes example.}
\label{table3}
\end{table}

\begin{figure}
\begin{center}
\resizebox{0.49\textwidth}{!}{% This file was created by matlab2tikz v0.2.3.
% Copyright (c) 2008--2012, Nico Schlömer <nico.schloemer@gmail.com>
% All rights reserved.
% 
% The latest updates can be retrieved from
%   http://www.mathworks.com/matlabcentral/fileexchange/22022-matlab2tikz
% where you can also make suggestions and rate matlab2tikz.
% 
% 
% 
\begin{tikzpicture}

\begin{semilogyaxis}[%
view={0}{90},
width=4.56842105263158in,
height=3.60315789473684in,
scale only axis,
xmin=0, xmax=25000,
xlabel={B},
ymin=0.1, ymax=10000,
yminorticks=true,
ylabel={IACT},
legend style={at={(0.97,0.97)},anchor=north east,fill=none,draw=none,align=left}]
\addplot [
color=black,
solid,
mark=x,
mark options={solid},
%forget plot
]
coordinates{
 (6,3.96236648665792)(4006,1.54622755555825)(8006,1.57917742989922)(12006,1.52553691150487)(16006,1.48495290610354)(20006,1.52569686683926) 
};
\addplot [
color=black,
solid,
mark=+,
mark options={solid},
%forget plot
]
coordinates{
 (6,39.6251380442732)(4006,1.55270354713852)(8006,1.53176742115167)(12006,1.53376463327228)(16006,1.49923944627164)(20006,1.55911336464772) 
};
\addplot [
color=black,
solid,
mark=asterisk,
mark options={solid},
%forget plot
]
coordinates{
 (6,6.38070483942407)(4006,2.26419858458264)(8006,2.31241821560676)(12006,2.27983272478649)(16006,2.23519483982961)(20006,2.20766773991467) 
};
\addplot [
color=blue,
dash pattern=on 1pt off 3pt on 3pt off 3pt,
mark=diamond,
mark options={solid},
%forget plot
]
coordinates{
 (6,235.270635007575)(4006,177.160454765154)(8006,163.53530564726)(12006,164.34798444784)(16006,161.574221358039)(20006,164.3682729996) 
};
\addplot [
color=blue,
dash pattern=on 1pt off 3pt on 3pt off 3pt,
mark=star,
mark options={solid},
%forget plot
]
coordinates{
 (6,7305.46559452605)(4006,182.036582598996)(8006,184.327097324045)(12006,181.825553082494)(16006,180.759723840751)(20006,157.446610163224) 
};
\addplot [
color=blue,
dash pattern=on 1pt off 3pt on 3pt off 3pt,
mark=star,
mark options={solid},
%forget plot
]
coordinates{
 (6,159.362175332837)(4006,182.787608309123)(8006,197.568113953789)(12006,187.622112608267)(16006,188.084479131735)(20006,151.979297191021) 
};
\addplot [
color=red,
dash pattern=on 1pt off 3pt on 3pt off 3pt,
mark=o,
mark options={solid},
%forget plot
]
coordinates{
 (6,0.991003281239685)(4006,0.998705234519927)(8006,0.998783985864447)(12006,0.996630719379699)(16006,0.997503153977436)(20006,1.00654434106416) 
};
\addplot [
color=red,
dash pattern=on 1pt off 3pt on 3pt off 3pt,
mark=square,
mark options={solid},
%forget plot
]
coordinates{
 (6,12.3160269761332)(4006,1.546638337845)(8006,1.55819701765434)(12006,1.56199371499433)(16006,1.54547266307102)(20006,1.54714775454662) 
};
\addplot [
color=red,
dash pattern=on 1pt off 3pt on 3pt off 3pt,
mark=triangle,
mark options={solid},
%forget plot
]
coordinates{
 (6,5.15742450330157)(4006,1.37266049277903)(8006,1.32837790767507)(12006,1.35570496890249)(16006,1.35573434909771)(20006,1.33060286854998) 
};
\addlegendentry{Gibbs $\theta$};
\addlegendentry{Gibbs $s_w$};
\addlegendentry{Gibbs $s_B$};
\addlegendentry{t-walk $\theta$};
\addlegendentry{t-walk $s_w$};
\addlegendentry{t-walk $s_B$};
\addlegendentry{tailor-made $\theta$};
\addlegendentry{tailor-made $s_w$};
\addlegendentry{tailor-made $s_B$};
\end{semilogyaxis}
\end{tikzpicture}%
}
\resizebox{0.49\textwidth}{!}{% This file was created by matlab2tikz v0.2.3.
% Copyright (c) 2008--2012, Nico Schlömer <nico.schloemer@gmail.com>
% All rights reserved.
% 
% The latest updates can be retrieved from
%   http://www.mathworks.com/matlabcentral/fileexchange/22022-matlab2tikz
% where you can also make suggestions and rate matlab2tikz.
% 
% 
% 
\begin{tikzpicture}

\begin{axis}[%
view={0}{90},
width=4.56842105263158in,
height=3.60315789473684in,
scale only axis,
xmin=0, xmax=25000,
xlabel={$B$},
ymin=0, ymax=0.002,
ylabel={CCES},
legend style={at={(0.97,0.97)},anchor=north east,fill=none,draw=none,align=left}]
\addplot [
color=black,
solid,
mark=x,
mark options={solid},
%forget plot
]
coordinates{
 (6,0.000604151599943329)(4006,0.000407720386574772)(8006,0.000563683134179625)(12006,0.000703923462803885)(16006,0.000921183313030378)(20006,0.00101633614381899) 
};
\addplot [
color=black,
solid,
mark=+,
mark options={solid},
%forget plot
]
coordinates{
 (6,0.00604174061839869)(4006,0.000409428022544052)(8006,0.00054676025913321)(12006,0.000707719953307539)(16006,0.000930045898739124)(20006,0.00103859639437121) 
};
\addplot [
color=black,
solid,
mark=asterisk,
mark options={solid},
%forget plot
]
coordinates{
 (6,0.000972881496571418)(4006,0.000597040143845313)(8006,0.000825411459553634)(12006,0.00105197569075024)(16006,0.0013865922477134)(20006,0.00147062799064844) 
};
\addplot [
color=red,
dash pattern=on 1pt off 3pt on 3pt off 3pt,
mark=o,
mark options={solid},
%forget plot
]
coordinates{
 (6,4.6312883373257e-05)(4006,5.06462499695034e-05)(8006,5.09797524253757e-05)(12006,4.72493853707585e-05)(16006,5.27547797288685e-05)(20006,5.49630985866206e-05) 
};
\addplot [
color=red,
dash pattern=on 1pt off 3pt on 3pt off 3pt,
mark=square,
mark options={solid},
%forget plot
]
coordinates{
 (6,0.000575568952964535)(4006,7.84329841913451e-05)(8006,7.95332116996497e-05)(12006,7.40527474734119e-05)(16006,8.17351500014846e-05)(20006,8.44831480263566e-05) 
};
\addplot [
color=red,
dash pattern=on 1pt off 3pt on 3pt off 3pt,
mark=triangle,
mark options={solid},
%forget plot
]
coordinates{
 (6,0.000241023621262878)(4006,6.9610235370366e-05)(8006,6.78028260555275e-05)(12006,6.42727795552943e-05)(16006,7.17004920458911e-05)(20006,7.26585542832944e-05) 
};
\addlegendentry{Gibbs $\theta$};
\addlegendentry{Gibbs $s_w$};
\addlegendentry{Gibbs $s_B$};
\addlegendentry{tailor-made $\theta$};
\addlegendentry{tailor-made $s_w$};
\addlegendentry{tailor-made $s_B$};
\end{axis}
\end{tikzpicture}%
}
\end{center}
\caption{IACT and CCES for increasing number of batches for the Dyes example.  CCES for t-walk is approximately $10^{-2}$ for this range of $B$.  Extrapolating linear growth of CCES for Gibbs $\theta$ and $s_w$ we expect t-walk to eventually beat Gibbs when $B > 2.6 \times 10^5$ (approximately).}
\label{fig5}
\end{figure}
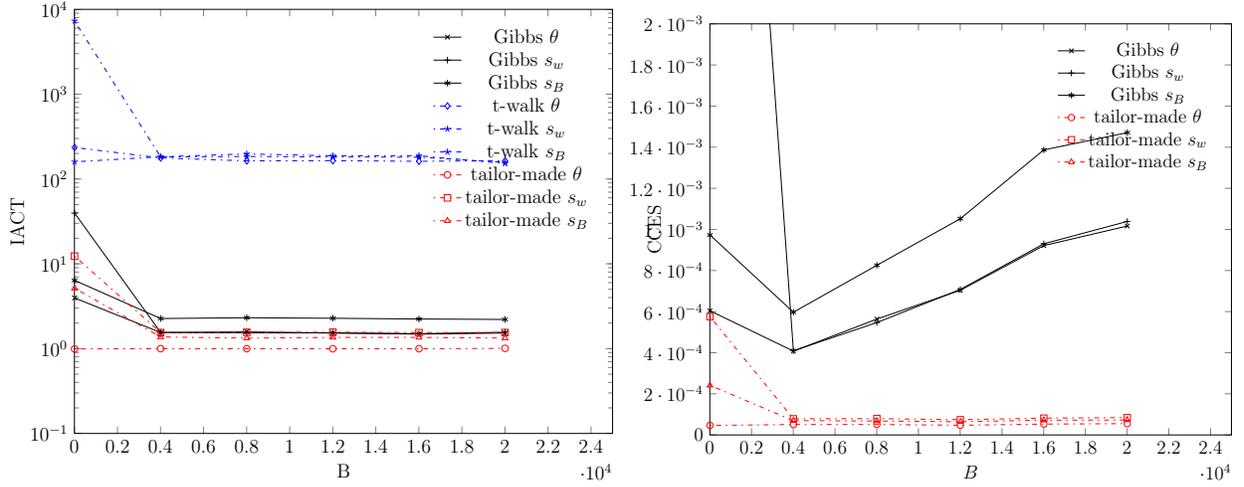

For this example, the main advantage of sampling from the marginal posterior distribution, rather than working with latent variables, is that the computational cost remains constant for increasing data size, in particular, as the number of batches increases.  Figure \ref{fig5} demonstrates this with larger data sets constructed by generating additional artificial batches of size $S$ using parameters $\theta = 1527$, $t_s = (3002)^{-1}$ and $t_b = (2264)^{-1}$.   We see that CCES grows linearly with $B$ for Gibbs sampling, whereas CCES remains independent of data size for both MCMC methods for sampling from $f_{\Theta,X,W|\mY}(\theta,x,w|\my)$.  IACT values appear to remain approximately constant for large $B$ for this problem.  The large initial drop in IACT for $s_w$ and $s_b$ could be attibuted to the extra data becoming informative about the values of $s_w$ and $s_b$.  

\subsection{Rat pups: a linear mixed model}
\label{sec:ratpup}

\cite{WWGbook}, model 3.1, used a linear mixed model to model weights of rat pups from litters with different treatments.  The model is
$$
	w_{ij} = \beta_0 + t_{1j} \beta_1 + t_{2j} \beta_2 + s_{ij} \beta_3 + l_j \beta_4  + t_{1j} s_{ij} \beta_5 + t_{2j} s_{ij} \beta_6 + u_{j} + \epsilon_{ij}
$$
where $w_{ij}$ is the weight of rat pup $i$ in litter $j$; $t_{1j} = 1$ if treatment to litter $j$ is `high', $0$ otherwise; $t_{2j} = 1$ if treatment to litter $j$ is `low', $0$ otherwise; $s_{ij} = 1$ if rat pup $i$ in litter $j$ is male, $0$ otherwise; $l_j$ is the size of litter $j$; $u_j \sim \normal(0,s_u)$ is the random effect on litter $j$; and $\epsilon_{ij} \sim \normal(0,s_\epsilon)$ is the error. 

We use data on $n = 322$ ratpups from $q=27$ litters from the webpage for \cite{WWGbook},~\url{http://www-personal.umich.edu/~bwest/almmussp.html}.
In matrix form the model is
$$
	\my = X \mbeta + Z \muu + \mepsilon
$$
where $\my \in \mathbb{R}^n$ is a vector of rat pup weights, $X \in \mathbb{R}^{n \times 7}$, $\mbeta = (\beta_0,\beta_1,\dotsc,\beta_6)^T$, $Z \in \mathbb{R}^{n \times q}$ and $\muu = (u_1,u_2,\dotsc,u_{q})^T$, and $\mepsilon \in \mathbb{R}^n$.  This linear effects model is used in many applications, particularly in plant and animal genetics applications where data sets may be very large, see for example \cite{Aguilar2011,VanRaden2008}.

Note that $Z^T Z$ is a diagonal matrix.  Then 
$$
	\my | \muu, \mbeta, s_\epsilon \sim \normal ( X \mbeta + Z \muu, s_\epsilon I ) \quad \mbox{and} \quad
	\muu | s_u \sim \normal(0,s_u I).
$$

From~\eqref{eqn:marg_posta} the posterior density function marginalized over $\muu$ is
\begin{multline}
\label{eq:rat1}
	f_{\mB,S_\epsilon,S_u|\mY}(\mbeta,s_\epsilon,s_u|\my) \propto \frac{1}{s_\epsilon^{n/2-q/2} s_u^{q/2}} \exp \left( -\frac{1}{2} g\left( \frac{s_\epsilon}{s_u} \right) - \frac{1}{2s_\epsilon} f\left(\mbeta, \frac{s_\epsilon}{s_u} \right) \right) \times \\ 
	f_{\mB}(\mbeta) f_{S_\epsilon}(s_\epsilon) f_{S_u}(s_u)
\end{multline}
where 
$$
	g(\lambda) = \log \det(Z^T Z + \lambda I) 
	\quad \mbox{and} \quad
	f(\mbeta,\lambda) = (X\mbeta-\my)^T (I-Z(Z^T Z + \lambda I)^{-1}Z^T) (X\mbeta - \my).
$$

We could also marginalize over $\mbeta$.  Instead we choose to retain the diagonal matrix structures in $f$ and $g$ in \eqref{eq:rat1}, that are also available in Gibbs sampling (see the expression for $\muu|\my,\mbeta,s_u,s_\epsilon$ later).

We perform the simplifying coordinate transformation $(s_\epsilon,s_u) \in [0,\infty)^2 \leftrightarrow (s,\lambda) \in [0,\infty)^2$ defined by
$$
	s = s_\epsilon \quad \mbox{and} \quad \lambda = \frac{s_\epsilon}{s_u},
$$
so that $\dd s_\epsilon \dd s_u = \frac{s}{\lambda^2} \, \dd s \dd \lambda$.
The density function in the new coordinate system satisfies
$$
	f_{\mB,S,\Lambda|\mY}(\mbeta,s,\lambda|\my) \propto \frac{\lambda^{q/2-2}}{s^{n/2-1}} \exp \left( -\frac{1}{2} g\left( \lambda \right) - \frac{1}{2s} f\left(\mbeta, \lambda \right) \right) f_{\mB}(\mbeta) f_{S_\epsilon}(s) f_{S_u}\left( \frac{s}{\lambda} \right).
$$

As for a similar model in \cite{HC1996}, we assume a uniform prior distribution on $\mbeta$ and hyperprior distributions on $s_\epsilon$ and $s_u$ such that  
$$
	f_{S_\epsilon}(s_\epsilon) \propto s_\epsilon^{-(a+1)} \quad \mbox{and} \quad
	f_{S_u}(s_u) \propto s_u^{-(b+1)}
$$
for some parameters $a,b \in (-1,\infty)$.  For our data, $\operatorname{Rank}(PZ)=23$, where $P = I - X (X^T X)^{-1} X^T$, so Theorem 1 of \cite{HC1996} implies that the joint posterior distribution $\mbeta,\muu,s_\epsilon,s_u|\my$ is proper if $a \in (-\frac{23}{2},0)$ and $a + b \in (-\frac{315}{2},\infty)$.  We use $a = b = -10^{-4}$ so that the posterior distribution is proper and the hyperprior distributions for $s_\epsilon$ and $s_u$ are close to Jeffrey's priors.

This choice of prior distribution allows us to apply both Gibbs sampling over the joint posterior distribution and our version of a MwG over the marginal posterior distribution.

\subsubsection[Metropolis within Gibbs]{Metropolis within Gibbs for $\mbeta,s,\lambda | \my$}

With our prior choice we obtain
\begin{align*}
	\mbeta | s,\lambda,\my &\sim \normal( (X^T W X)^{-1} X^T W \my, s (X^T W X)^{-1}), \\
	s | \mbeta,\lambda,\my &\sim \operatorname{InvGa}\left( \frac{n}{2} + a + b, \frac{1}{2 f(\mbeta,\lambda)} \right),
\end{align*}
where $W = I - Z(Z^TZ + \lambda I)^{-1} Z^T$, and $\operatorname{InvGa}(r,s)$ is an inverse gamma distribution with probability density function $\pi(t) \propto t^{-(r+1)} \exp(-\frac{1}{st})$ for $t > 0$.  

In particular, to draw a sample from $\mbeta|s,\lambda,\my$ we first construct the matrix $X^T W X$ and vector $X^T W y$, compute a Cholesky factorization of $X^T W X = L L^T$, compute $\mz = X^T W y + \sqrt{s} L \mxi$ where $\mxi \sim \normal(0,I)$, then solve $(X^T W X) \mbeta = \mz$ for $\mbeta$.

To draw a sample from $s | \mbeta, \lambda,\my$ we compute $f(\mbeta,\lambda)$, sample $z \sim \operatorname{Ga}(\frac{n}{2}+a+b,\frac{1}{2} f(\mbeta,\lambda))$, then define $s = z^{-1}$.

To obtain a sample from $\lambda|\mbeta,s,\my$ we perform one iteration of RWM with proposal $\lambda' \sim \normal(\lambda,w^2)$ for some $w$, then accept or reject according to Metropolis.  If $\lambda' <0$ then we always reject.  The target density function for $\lambda|\mbeta,s,\my$ is
$$
	\pi(\lambda|\mbeta,s,\my) \propto \lambda^{q/2+a-1} \exp \left( -\frac{1}{2} g(\lambda) - \frac{1}{2s}f(\mbeta,\lambda) \right).
$$

Thus, efficiency of our MwG sampler depends on our ability to efficiently assemble $X^T W X$, $X^T W y$, as well as evaluating $f(\mbeta,\lambda)$ and $g(\lambda)$.  Since $f(\mbeta,\lambda) = \beta^T X^T W X \beta - 2 \beta^T X^T W y + y^T W y$, we can reduce our computing list to $X^T W X$, $X^T W y$, $y^T W y$ and $g(\lambda)$.  In the appendix we show how to reduce the number of operations requried to compute all of these quantities so that they depend only on the maximum litter size.  Note that the expected maximum litter size is a very slowly increasing function of $q$ since we assume litter size is Poisson distributed with a finite mean, so the cost of MwG is almost sample size independent.

\subsubsection{Simulations}

We compare our calculations using MwG sampling for $\mbeta,s,\lambda|\my$ with the usual Gibbs sampling method for $\mbeta,\mmu,s_u,s_\epsilon|\my$, which iteratively draws independent samples from
\begin{align*}
	s_u | \my, \mbeta, \muu, s_\epsilon &\sim \operatorname{InvGa}\left( a + \frac{q}{2}, \frac{2}{|\muu|^2} \right), \\
	s_\epsilon | \my, \mbeta, \muu, s_u &\sim \operatorname{InvGa}\left( b + \frac{n}{2}, \frac{2}{|\my - X \mbeta - Z \muu|^2} \right), \\
	\muu | \my, \mbeta, s_u, s_\epsilon &\sim \normal\left( \left(Z^T Z + \frac{s_\epsilon}{s_u} I \right)^{-1} Z^T(\my - X \mbeta) , s_\epsilon \left(Z^T Z + \frac{s_\epsilon}{s_u} I \right)^{-1} \right), \\
	\mbeta | \my, \muu, s_u, s_\epsilon &\sim \normal \left( (X^T X)^{-1} X^T (\my - Z \muu), s_\epsilon (X^T X)^{-1} \right).
\end{align*}

In our implementation we use the fact that $z \sim \operatorname{Ga}(r,s^{-1})$ implies $z^{-1} \sim \operatorname{InvGa}(r,s)$, and also, $\mz \sim \normal(\mb,A)$ implies $A^{-1} \mz \sim \normal(A^{-1}\mb,A^{-1})$.  Note that $Z^T Z$ is a diagonal matrix for this rat pup model, and we pre-compute $(X^T X)^{-1}$ and a Cholesky factorization of $X^T X$ to efficiently compute the action of $(X^T X)^{-1}$ and samples from $\normal(0,X^T X)$.

Chain length was $10^5$ after a burn in of $10^3$.  We used initial $s_\epsilon = 0.16$ and $s_u = 0.105$, and for Gibbs we also used initial $\mbeta = (X^TX)^{-1} X^T y$ and $\mmu = \left(Z^T Z + \frac{s_\epsilon}{s_u} I \right)^{-1} Z^T(\my - X \mbeta)$.

For MwG we computed a training run of length $10^4$, then choose $w$ to be twice the standard deviation of the $\lambda$ values from the training run.  This ensured that approximately half of the subsequent $\lambda$ proposals were accepted.

Using $\mbeta = (7.9103, -0.7994,-0.3810,0.4115,-0.1281,-0.1078,-0.0842)^T$, $s_u = 0.1055$ and $s_\epsilon = 0.1648$ (mean values from the previous Gibbs chain) and $l$ equal to the mean of the rat pup litter sizes for the genuine data, we generated additional artificial data.  For each additional rat pup litter we choose $l_j \sim \operatorname{Po}(l)$, treatment uniformly, and $u_j \sim \normal(0,s_u)$.  We then choose each rat pup within a litter to have uniformly random gender, and weight given by the model with $\epsilon_{ij} \sim \normal(0,s_\epsilon)$.

\begin{figure}
\begin{center}
\resizebox{0.49\textwidth}{!}{% This file was created by matlab2tikz v0.2.3.
% Copyright (c) 2008--2012, Nico Schlömer <nico.schloemer@gmail.com>
% All rights reserved.
% 
% The latest updates can be retrieved from
%   http://www.mathworks.com/matlabcentral/fileexchange/22022-matlab2tikz
% where you can also make suggestions and rate matlab2tikz.
% 
% 
% 
\begin{tikzpicture}

\begin{axis}[%
view={0}{90},
width=4.56842105263158in,
height=3.60315789473684in,
scale only axis,
%every outer x axis line/.append style={darkgray!60!black},
%every x tick label/.append style={font=\color{darkgray!60!black}},
xmin=0, xmax=20000,
xlabel={$q$},
%every outer y axis line/.append style={darkgray!60!black},
%every y tick label/.append style={font=\color{darkgray!60!black}},
ymin=1, ymax=5,
ylabel={IACT},
legend style={at={(0.97,0.5)},anchor=east,align=left,fill=none,draw=none}]
\addplot [
color=black,
solid,
mark=x,
mark options={solid},
%forget plot
]
coordinates{
 (1000,1.62957544022427)(4800,1.66495876651328)(8600,1.56208578694168)(12400,1.59743269489505)(16200,1.63225787228505)(20000,1.61698403625829) 
};
\addplot [
color=black,
solid,
mark=+,
mark options={solid},
%forget plot
]
coordinates{
 (1000,1.18063598315335)(4800,1.18065808588653)(8600,1.20070780886216)(12400,1.19110718519621)(16200,1.19760930927522)(20000,1.19025385455785) 
};
\addplot [
color=red,
dashed,
mark=o,
mark options={solid},
%forget plot
]
coordinates{
 (1000,4.40934247761813)(4800,4.55356943335842)(8600,4.48973075979316)(12400,4.71492814508257)(16200,4.6834085394757)(20000,4.42358906786722) 
};
\addplot [
color=red,
dashed,
mark=square,
mark options={solid},
%forget plot
]
coordinates{
 (1000,1.44068684834995)(4800,1.4463073964506)(8600,1.4958855464044)(12400,1.46781498399624)(16200,1.47929153977948)(20000,1.42081576792966) 
};
\addlegendentry{Gibbs $s_u$};
\addlegendentry{Gibbs $s_\epsilon$};
\addlegendentry{tailor-made $s_u$};
\addlegendentry{tailor-made $s_\epsilon$};
\end{axis}
\end{tikzpicture}%
}
\resizebox{0.49\textwidth}{!}{% This file was created by matlab2tikz v0.2.3.
% Copyright (c) 2008--2012, Nico Schlömer <nico.schloemer@gmail.com>
% All rights reserved.
% 
% The latest updates can be retrieved from
%   http://www.mathworks.com/matlabcentral/fileexchange/22022-matlab2tikz
% where you can also make suggestions and rate matlab2tikz.
% 
% 
% 
\begin{tikzpicture}

\begin{axis}[%
view={0}{90},
width=4.56842105263158in,
height=3.60315789473684in,
scale only axis,
%every outer x axis line/.append style={darkgray!60!black},
%every x tick label/.append style={font=\color{darkgray!60!black}},
xmin=0, xmax=20000,
xlabel={$q$},
%every outer y axis line/.append style={darkgray!60!black},
%every y tick label/.append style={font=\color{darkgray!60!black}},
ymin=0, ymax=0.003,
ylabel={CCES},
legend style={at={(0.03,0.97)},anchor=north west,align=left,draw=none,fill=none}]
\addplot [
color=black,
solid,
mark=x,
mark options={solid},
%forget plot
]
coordinates{
 (1000,0.000321829984432367)(4800,0.00079176736640987)(8600,0.00105677152757562)(12400,0.0016169764614765)(16200,0.00204717415487993)(20000,0.00286444242497084) 
};
\addplot [
color=black,
solid,
mark=+,
mark options={solid},
%forget plot
]
coordinates{
 (1000,0.000233167517562882)(4800,0.000561459275805701)(8600,0.00081229458455512)(12400,0.00120567976836378)(16200,0.00150203890403642)(20000,0.00210850172978205) 
};
\addplot [
color=red,
dashed,
mark=o,
mark options={solid},
%forget plot
]
coordinates{
 (1000,0.000964735888116877)(4800,0.00101283345207502)(8600,0.00107571032902289)(12400,0.00118107754963422)(16200,0.00108952469921055)(20000,0.00112178447760927) 
};
\addplot [
color=red,
dashed,
mark=square,
mark options={solid},
%forget plot
]
coordinates{
 (1000,0.000315213053464603)(4800,0.000321696755599552)(8600,0.000358404461067817)(12400,0.000367683933088721)(16200,0.000344134972710149)(20000,0.000360306766644228) 
};
\addlegendentry{Gibbs $s_u$};
\addlegendentry{Gibbs $s_\epsilon$};
\addlegendentry{tailor-made $s_u$};
\addlegendentry{tailor-made $s_\epsilon$};
\end{axis}
\end{tikzpicture}%
}
\end{center}
\caption{Rat pup example; IACT and CCES for increasing number of rat pup litters.}
\label{fig:ratpup2}
\end{figure}
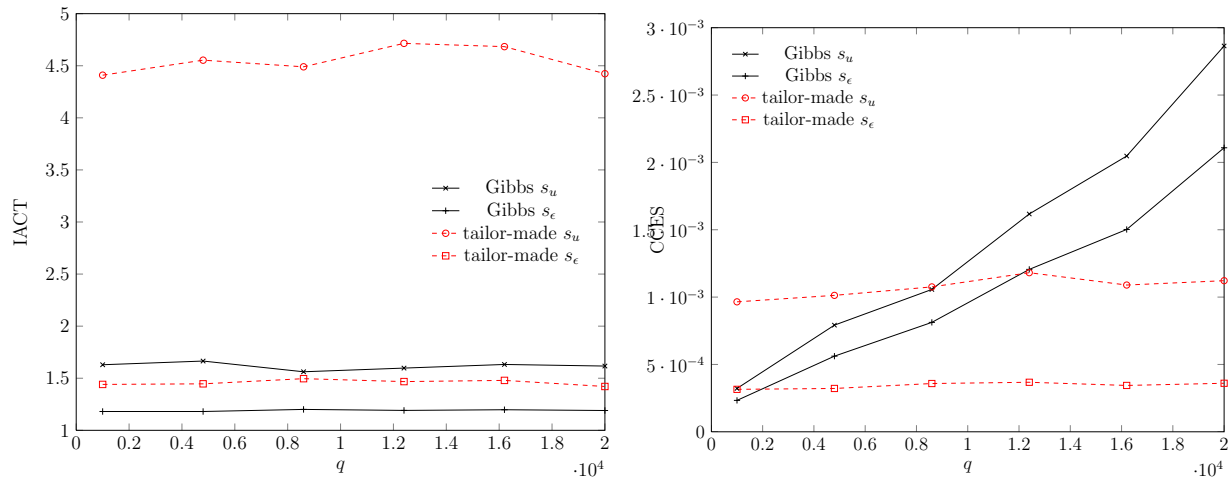

In Figure \ref{fig:ratpup2}, as expected, CCES for our MwG sampler remains constant, while it grows linearly for Gibbs.  When calculating CCES we ignored the setup cost because it is similar for both algorithms (slightly more for MwG given the additional precomputing steps and training run).  

However, Figure \ref{fig:ratpup2} also shows that data size needs to grow substantially before our MwG sampler outperforms Gibbs sampling.  In part, this is due to the already good performance of the Gibbs sampler for this example; since $Z^T Z$ is a diagonal matrix the Gibbs sampler only costs order $q$ operations per iteration.  In other mixed linear models, if $Z^T Z$ or the covariance matrices $s_u I$ or $s_\epsilon I$ are not diagonal \citep{VanRaden2008} then we would expect the cost per iteration to be order $q^2$ per iteration and marginalizing over $\muu$ would lead to greater efficiency gains.

\section{Discussion}

In hierarchical models for which the full conditional posterior distribution over latent variables belongs to a known family of distributions, we can evaluate the density function for the posterior distribution over the hyperparameters allowing us to perform MCMC on a low-dimensional state space.  This should provide efficiency gains since correlations between the hyperparameters and latent variables, that can adversely effect MCMC on the joint posterior distribution, are irrelevant.

For large data sizes, the computational bottleneck becomes the cost of evaluating the marginal posterior density function, whatever MCMC method is used.  If we are lucky, then we can also derive a low-dimensional, sample size independent, sufficient statistic for the marginal posterior distribution that can be precomputed.  The `on-line' cost of evaluating the marginal posterior density function is then dependent on the complexity of the sufficient statistic, rather than data size.

Armed with a cheap way of computing the marginal posterior density function, the user is then free to apply whatever MCMC they wish to use to draw samples from the posterior distribution over hyperparameters.  A good `no-think' method is t-walk because it requires very little additional input from the user.  However, for the price of a little bit more work, efficient bespoke methods can also be implemented.  

If the user is interested in quantities that depend on the latent variables then they should still perform MCMC over the hyperparameters first.  Once independent hyperparameter samples have been obtained, the user can then draw independent samples from the full conditional posterior distribution over the latent variables.  The result is an independent sample from the joint posterior distribution.
This technique could be called marginal then conditional (MTC) sampling.

Our approach challenges the conventional approach of applying a Gibbs sampler in situations where full conditional posterior distributions over all variables are available, especially in high dimensions when the cost of Gibbs grows with the number of latent variables.  

We can also avoid choosing conjugate prior distributions simply because they make Gibbs sampling possible.  Our approach lets the prior choice of hyperparameters be determined from modeling considerations rather than computational convenience.

Even in the case when a fixed-dimensional sufficient statistic cannot be found for the marginal posterior distribution over hyperparameters, we still think that efficient MCMC is possible.  Our intuition originates from the observation that the marginal posterior density function tends to be a slowly varying function of the hyperparameters, and it should be possible to approximate that function efficiently.

\bibliographystyle{chicago}
\bibliography{NCF}

\appendix
\section{Dyes example: coordinate transformation}

The simplifying coordinate transformation used in Section \ref{sec:dyes} is the result of the two coordinate transformations that first eliminate $S t_w$ and $t_b + S t_w$ terms, and then eliminate $\frac{v-x}{v}$ terms.  The extracting of two conditional distributions in Section \ref{sec:dyesmwg} relies on the choice of conjugate priors.

First, perform the coordinate transformation $(t_w,t_b) \in [0,\infty)^2 \leftrightarrow (x,v) \in [0,\infty)^2$ defined by $x = S t_w$ and $v = t_b + S t_w$.  Then $\dd t_w \dd t_b = S^{-1} \, \dd x \dd v$ and \eqref{eq:f1} becomes
\begin{multline*}
	f_{\Theta,X,V|\mY}(\theta,x,v|\my) \propto x^{BS/2} \left( \frac{v-x}{v} \right)^{B/2} \times \\ 
	\exp \left( \frac{B}{2} \left[ r_1 \frac{x^2}{v} - r_3 x - x \left( \frac{v-x}{v} \right) \left( (\theta-r_2)^2 - r_2^2 \right) \right]\right) f_{\Theta}(\theta) f_{T_w}(\tsfrac{x}{S}) f_{T_b}(v-x).
\end{multline*}
Next, perform the coordinate transformation $(x,v) \in [0,\infty)^2 \leftrightarrow (x,w) \in [0,\infty)\times [0,1]$ defined by $x = x$ and $w = \frac{v-x}{v}$.  Then $\dd x \dd v = \frac{x}{(1-w)^2} \dd x \dd w$ and the new density function is
$$
	f_{\Theta,X,W|\mY}(\theta,x,w|\my) \propto 
	\frac{x^{BS/2} w^{B/2}}{(1-w)^2} \exp \left( - \frac{B}{2} \left[ R_3 x + \left( (\theta-r_2)^2 + R_2\right) xw  \right] \right) f_{\Theta}(\theta) f_{T_w}\left( \tsfrac{x}{S} \right) f_{T_B}\left( \tsfrac{xw}{1-w} \right)
$$
where $R_2 = r_1 - r_2^2$ and $R_3 = r_3 - r_1$.

\section[Rat pup example: fast on-line computation]{Rat pup example: fast on-line computation of $X^T W X$, $X^T W y$, $y^T W y$ and $g(\lambda)$}

In the rat pup example, the matrix $Z^T Z$ is diagonal with the size of each rat pup litter on the diagonal.  Not all rat pup litter sizes are different.  We exploit this to precompute certain terms so that the on-line cost of computing $X^T W X$, $X^T W y$, $y^T W y$ and $g(\lambda)$ is independent of $q$.

Let $\mathcal{J}(i) = \{ j=1,2,\dotsc,q : (Z^T Z)_{jj} = i \}$, so that $\mathcal{J}(i)$ is the set of litters with size $i$, let $r$ be the maximum litter size, and let $\mc_j$ denote the $j^{\mathrm{th}}$ column of $X^T Z$.  Then
\begin{align*}
	X^T W X &= X^T (I - Z(Z^T Z + \lambda I)^{-1} Z^T) X \\
	 &= X^T X - \sum_{j=1}^q \frac{1}{(Z^T Z)_{jj} + \lambda} \mc_j \mc_j^T \\
	 &= X^T X - \sum_{i=1}^{r} \frac{1}{i+\lambda} \sum_{j \in \mathcal{J}(i)} \mc_j \mc_j^T.
\end{align*}
Therefore, we achieve fast on-line computation of $X^T W X$ if we precompute the $p \times p$ matrices $X^T X$ and $\sum_{j \in \mathcal{J}(i)} \mc_j \mc_j^T$ for each $i=1,\dotsc,r$.

Similar tricks work for $X^T W y$, $y^T W y$ and $g(\lambda)$.

\end{document}